\documentclass[preprint,showpacs,preprintnumbers,amsmath,amssymb]{revtex4}
\usepackage{graphicx}
\usepackage{dcolumn}
\usepackage{bm}
\newcommand{\be}{\begin{eqnarray}}
\newcommand{\ee}{\end{eqnarray}}

\begin{document}
\title{Zero modes, energy gap, and edge states of 
anisotropic honeycomb lattice in a magnetic field
}
\author{Kenta Esaki$^1$, Masatoshi Sato$^1$, Mahito Kohmoto$^1$,
and Bertrand I. Halperin$^2$}
\affiliation{%
$^1$Institute for Solid State Physics,
Kashiwanoha 5-1-5, Kashiwa, Chiba 277-8581, Japan\\
$^2$Physics Department, Harvard University,
 Cambridge, Massachusetts 02138
}%
\affiliation{%
}%
\date{\today}
\begin{abstract}
We present systematic study of zero modes and gaps by introducing effects of
anisotropy of hopping integrals for a tight-binding model 
on the honeycomb lattice in a magnetic field.
The condition for the existence of zero modes is analytically derived.
From the condition, it is found that a tiny anisotropy for graphene 
is sufficient to open a gap around zero energy in a magnetic field.
This gap behaves as a non-perturbative and exponential form 
as a function of the magnetic field.
The non-analytic behavior with respect to the magnetic field
can be understood as tunneling effects between energy levels
around two Dirac zero modes appearing in the honeycomb lattice,
and an explicit form of the gap around zero energy
is obtained by the WKB method near the merging point of 
these Dirac zero modes.
Effects of the anisotropy for the honeycomb lattices with boundaries
are also studied.
The condition for the existence of zero energy edge states in a magnetic field
is analytically derived. On the basis of the condition, 
it is recognized that anisotropy of the hopping integrals induces
abrupt changes of the number of zero energy edge states,
which depend on the shapes of the edges sensitively.

\end{abstract}


\maketitle

\section{Introduction}

\begin{figure}
 \begin{center}
  \includegraphics[width=8.0cm,clip]{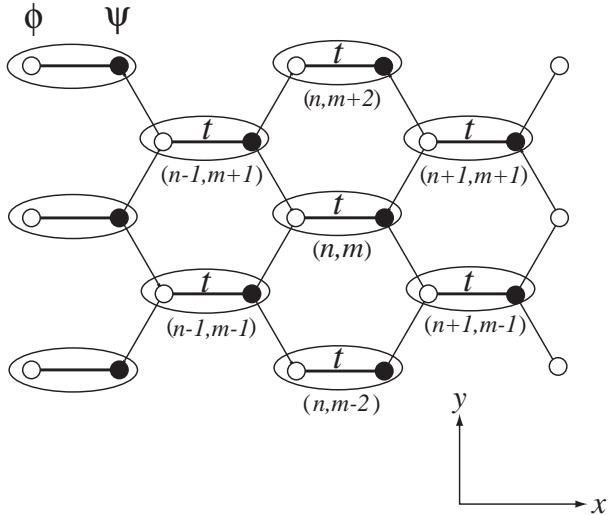}
\caption{\label{fig:honeycomb_lattice}
 The honeycomb lattice. 
 The hopping integrals of the horizontal bonds are $t$, 
 and those for the other bonds are 1.
 A magnetic flux $2\pi\Phi$ is applied through the unit hexagon.}
\end{center}
\end{figure} 

Recent experiments on graphene\cite{KS05,YZ05,ML06,ZJ07,RS07} 
have led to renewed interest
in physical properties of electrons on the honeycomb lattice.
Despite its simple structure, the honeycomb lattice provides non-trivial
physical phenomena which can not be observed in the ordinary square lattice.
Among them, much attention has been paid to its peculiar dispersion.
In the absence of a magnetic field, the honeycomb lattice has 
$E=0$ zero modes at the corners K and K' of the Brillouin zone.
By treating these zero modes as Dirac fermions, the unconventional
quantization of the Hall conductance observed for graphene 
was explained\cite{YZ02,VP05,NM06}, 
although the full proper theoretical treatment of the Hall
conductance on the honeycomb lattice was made very recently\cite{MS08}.
Moreover, when the system has a boundary, there are $E=0$ edge modes
localized on the boundary.
The existence of the $E=0$ edge modes depends on a choice of the
boundary, and for zigzag and bearded edges there occur large density of
states localized on these edges at the Fermi energy due to their flat
band structures\cite{Klein,MF96,KN96,KW99,KK02,SR02,ME06,
BA06,BA07,YH06,YH07}.
In addition, it is suggested that the $E=0$ edge modes induce
charge accumulation on these edges\cite{MA08,MA08_2}.

In this paper, we study properties of these $E=0$ zero modes 
in the presence of anisotropy of the hopping integrals 
in the honeycomb lattice.
Recently, the anisotropy of the hopping integrals was
introduced by replacing one of the hopping integrals 
with a general value $t$\cite{MK06,MK06_2,PD08,MK07,MS08,MK07_edge,VM08} 
in order to investigate the unconventional quantum Hall effects on graphene
 (see Fig.\ref{fig:honeycomb_lattice}). 
(For $t\ne 1$, we have the anisotropic honeycomb lattice.)
In Ref.\cite{MS08}, by using topological arguments, 
an algebraic expression of the quantum Hall conductance was obtained for
almost all gaps including subband gaps, and it was shown that
the unconventional quantization of the Hall conductance 
in a weak magnetic field is
realized for weak $t$ ($0<t\lesssim 1$), 
while only the conventional quantization
is obtained for strong $t$ ($t>2$).
Furthermore, for the graphene case ($t=1$), the unconventional quantization 
was found to persist up to the Van Hove singularity\cite{MS08,YH06}.

The anisotropy of the hopping parameters is also known to change the peculiar
dispersion mentioned above\cite{MK06_2,PD08,MK07,VM08}. 
However, in the absence of a magnetic field,
its influence is restrictive:
Although a gap opens around zero energy for $t>2$,  
there remain two $E=0$ zero modes in the Brillouin zone 
for $0<t<2$\cite{MK06_2}. 
Therefore, a large anisotropy is needed to change the zero mode structure.
As well as the zero mode structure, that of zero energy edge states 
was shown to change by a large anisotropy\cite{MK07_edge}.

In this paper, it will be shown the situation is drastically changed in the
presence of a magnetic field. 
We analytically derive the condition 
for the existence of zero modes in a magnetic flux
$2\pi\Phi=2\pi p/q$ ($p$ and $q$ are mutually prime integers),
and  from the condition it is found that, 
in the limit of $q\rightarrow \infty$, 
zero modes exist only for $0<t\le 1$, but
a gap around zero energy opens for $t> 1$.
In other words, a small anisotropy $t=1+\epsilon$ ($0<\epsilon\ll 1$) is
sufficient to open a gap in the presence of a weak magnetic field.

For $1<t<2$, the gap around zero energy in a weak magnetic field 
behaves as a non-perturbative and exponential form 
as a function of $\Phi$.
It will be shown that this behavior is naturally explained in terms of
the spontaneous breaking of supersymmetry \cite{Witten81,Witten81_2}.
In particular, an explicit form of the gap around zero energy
for $t\sim 2$ is obtained by the WKB method.
At $t=2$, the gap around zero energy in a weak magnetic field
is found to make a transition from an exponential (non-perturbative) 
to a power-law (perturbative) behavior 
as a function of $\Phi$, and 
for $t>2$, the energy bands in a weak magnetic field 
show linear dependence on $\Phi$.

We will also show that the structure of $E=0$ edge states in the presence of a
magnetic field is different from that in the absence of a magnetic field.  
The condition for the existence of zero energy edge states in a magnetic field
is analytically derived, and it is found that the anisotropy of the 
hopping integrals induces abrupt changes of the number of 
zero energy edge states, which also sensitively depend on shapes of the edges.

The organization of this paper is as follows. 
In Sec.\ref{sec:model}, we present our model.
The condition for the existence of zero modes
in a magnetic field is analytically derived in Sec.\ref{sec:condition},
both from the secular equation and
from the normalizability condition of states with zero energy.
On the basis of the condition for the existence of zero modes,
the energy spectrum near zero energy in a weak magnetic field 
is systematically examined in Sec.\ref{sec:small_magnetic}.
In Sec.\ref{sec:edge}, zero energy edge states are analyzed,
where crucial roles of the anisotropy of the hopping integrals are
recognized again.
Finally, we summarize our results and discuss 
possible experimental realization 
of anisotropy of the hopping integrals in Sec.\ref{sec:summary}.

\section{Tight-binding model on the honeycomb lattice
in a magnetic field}
\label{sec:model}

Let us consider the tight-binding model on the honeycomb lattice
 with nearest-neighbor hopping in a magnetic field
as shown in Fig.\ref{fig:honeycomb_lattice}.
By denoting wave functions on two sublattices of the honeycomb lattice 
as $\psi_{n,m}$ and $\phi_{n,m}$, respectively, 
the tight-binding model is given by
\be
E\psi_{n,m}&=& \phi_{n+1,m-1}+e^{2 i\pi\Phi n}\phi_{n+1,m+1}+t\phi_{n,m},
 \nonumber\\
E\phi_{n,m}&=&\psi_{n-1,m+1}+e^{-2 i\pi\Phi (n-1)}\psi_{n-1,m-1}+t\psi_{n,m},
\label{magnetic2}
\ee
where a magnetic flux through the unit hexagon is given by $2\pi\Phi$. 
Here we have introduced anisotropy of the hopping integrals:
The hopping integrals of the
horizontal bonds are $t$, and those for the other bonds are 1.
For simplicity, we neglect the spin degrees of freedom in the following.

\section{The condition for the existence of zero modes}
\label{sec:condition}

For the isotropic case ($t=1$), it was found that
zero modes exist for all (rational) values of $\Phi$\cite{Rammal}.
We now derive the condition for the existence of zero modes
in the anisotropic case.

Before examining $\Phi\ne 0$, let us first consider $\Phi=0$\cite{MK06_2}.
For $\Phi=0$, (\ref{magnetic2}) gives
\be
E\psi_{n,m}&=&  \phi_{n+1,m-1}+\phi_{n+1,m+1}+t\phi_{n,m}, \nonumber\\
E\phi_{n,m}&=&  \psi_{n-1,m+1}+\psi_{n-1,m-1}+t\psi_{n,m}.
\label{non_magnetic}
\ee
From the Bloch's theorem, the wave functions are written as
\footnote{Note that, from the definition (\ref{Bloch_uv}), $k_x$ and $k_y$
have different units of length from each other in our convention.
See Eqs. (\ref{kq1}) and (\ref{kq2}) in Appendix \ref{sec:appendixA}.} 
\be
\psi_{n,m}=e^{i k_x n+i k_y m} \psi({\bm k}),\quad
\phi_{n,m}=e^{i k_x n+i k_y m} \phi({\bm k}).
\label{Bloch_uv}
\ee
Substituting (\ref{Bloch_uv}) into (\ref{non_magnetic}), we have
\be
{\cal Q}({\bm k})\left(
\begin{array}{c} 
  {\psi({\bm k})}\\ 
  {\phi({\bm k})}
\end{array} \right)
=E\left(
\begin{array}{c} 
  {\psi({\bm k})}\\ 
  {\phi({\bm k})}
\end{array} \right),
\ee
where ${\cal Q}({\bm k})$ is given by
\be
{\cal Q}({\bm k})=\left(
\begin{array} {cc}
0 &  {\cal D}({\bm k}) \\
 {\cal D}^*({\bm k}) & 0
\end{array} 
\right),
\quad
{\cal D} ({\bm k})=t+2e^{i k_x}\cos k_y.
\label{delta_zero}
\ee
The eigenenergies $E$ are given by
\be
E&=&\pm |{\cal D}({\bm k})| \nonumber\\
 &=&\pm \sqrt{(t+2\cos k_x\cos k_y)^2+4 \sin^2 k_x \cos^2 k_y}.
\label{eigen_energy}
\ee
From (\ref{eigen_energy}) with $E=0$,
we find two zero modes at
\be
{\bm k}_{0}^{+}: (k_x^{0+},k_y^{0+})=\left(\pi,\cos^{-1}\frac{t}{2} \right),
\quad
{\bm k}_{0}^{-}: (k_x^{0-},k_y^{0-})=\left(\pi,-\cos^{-1}\frac{t}{2} \right),
\label{zero_modes}
\ee
for $0<t<2$.
By expanding $k_x$ and $k_y$ around $k_x^{0\pm}$ and $k_y^{0\pm}$ 
in (\ref{zero_modes}),
\be
k_x=k_x^{0\pm}+p_x,
\quad
k_y=k_y^{0\pm}+p_y,
\quad
(|p_x|,|p_y|\ll 1),
\ee
${\cal D}({\bm k})$ is given by
\be
{\cal D}_{\pm}({\bm p})=-itp_x\pm\sqrt{4-t^2}p_y,
\label{delta_1}
\ee
where ${\cal D}_{+}({\bm p})$ and ${\cal D}_{-}({\bm p})$ are those
near ${\bm k}_{0}^{+}$ and ${\bm k}_{0}^{-}$, respectively.
From (\ref{eigen_energy}) and (\ref{delta_1}),
the dispersion relation of the Dirac zero mode is obtained:
\be
E=\pm \sqrt{t^2 p_x^2+(4-t^2) p_y^2}.
\ee
For $t=2$, the two Dirac zero modes merge into a confluent point 
\be 
(k_x^*,k_y^*)=(\pi,0),
\label{confluent}
\ee
and for $t>2$, we have a gap around zero energy.

\subsection{Derivation from a secular equation}
\label{sec:zero_mode}

Now we consider $\Phi\ne 0$.
We suppose that $\Phi$ is a rational number, 
$\Phi=p/q$ ($p$ and $q$ are mutually 
prime integers).
Since Eq.(\ref{magnetic2}) has translational symmetry in the $y$-direction, 
the wave functions are written as
\be
\psi_{n,m}=e^{i k m} \psi_n,\quad \phi_{n,m}=e^{i k m} \phi_n,
\label{translation}
\ee
and (\ref{magnetic2}) becomes
\be
E\psi_n&=&(e^{-ik}+e^{ik+2i\pi\Phi n})\phi_{n+1}+t\phi_n, \nonumber \\
E\phi_n&=&(e^{ik}+e^{-ik-2i\pi\Phi (n-1)})\psi_{n-1}+t\psi_n.
\label{magnetic3}
\ee
By the gauge transformation $\psi_n \to e^{i k n}\psi_n$ and
$\phi_n \to e^{i k n}\phi_n$, (\ref{magnetic3}) is rewritten as
\be
E\psi_n&=&A_n \phi_{n+1}+t\phi_n, \nonumber \\
E\phi_n&=&A_{n-1}^* \psi_{n-1}+t\psi_n,
\label{magnetic4}
\ee
where $A_n=1+\exp[i(\theta_1+2\pi\Phi n)]$ with $\theta_1=2 k$.
Since the spectrum is found to be invariant under the transformation
$\theta_1 \to \theta_1+2\pi/q$,
we can restrict the range of $\theta_1$ to $0\le \theta_1 \le 2\pi/q$
without loss of generality.
Moreover, in (\ref{magnetic4}), ($\psi_n$,$\phi_n$) and ($\psi_{n+q}$,$\phi_{n+q}$)
obey the same equation, thus from the Bloch's theorem, we have
\be
\psi_{n+q}=\exp(i q \theta_2) \psi_n,\quad
\phi_{n+q}=\exp(i q \theta_2) \phi_n,
\label{Bloch_g}
\ee
where $\theta_2$ satisfies $0\le \theta_2 \le 2\pi/q$.
Therefore (\ref{magnetic4}) reduces to the eigenequation 
of a $2q\times 2q$ matrix.
In the secular equation of this, all non-constant terms
containing less than $q$ factors  
of $e^{i\theta_1}$ should cancel out each other
since the eigenvalue has periodicity $2\pi/q$ with respect to $\theta_1$.
From this property, it is found that the secular equation is written as
the following form:
\be
F(E^2)+f(\theta_1,\theta_2)=0,
\label{secular_equation}
\ee
where $F(E^2)$ is a $q$th-order polynomial of $E^2$ with $F(0)=0$,
and it is independent of $(\theta_1,\theta_2)$.
The secular determinant for (\ref{magnetic4}) with $E=0$ 
determines $f(\theta_1,\theta_2)$ as  
\be
f(\theta_1,\theta_2)&=&
\left|\det\begin{pmatrix}
   t &  A_1  &       &           &          \\
     &  t    &  A_2  &           &          \\
     &       &\ddots & \ddots    &          \\
     &       &       &  t        &  A_{q-1} \\
 e^{i q \theta_2} A_q    &      &       &           &     t 
  \end{pmatrix}
\right|^2
=\left|t^q+(-1)^{q-1}e^{i q \theta_2}\prod_{n=1}^q A_n\right|^2\nonumber\\
&=&\left|t^q+(-1)^{q+1}2\cos\left(\frac{q}{2}\theta_1+\frac{q+1}{2}\pi \right)
e^{i(q \theta_2+\frac{q}{2}\theta_1+ \frac{q+1}{2}\pi)}\right|^2.
\label{f_expression}
\ee
Here we have used
\be
\prod_{n=1}^q A_n 
&=&\prod_{n=1}^q [1+e^{i(\theta_1+2\pi\Phi n)}] 
=1+e^{i q \theta_1} \prod_{n=1}^q e^{ i 2\pi\Phi n} \nonumber\\
&=& 1+ (-1)^{p(q+1)} e^{i q \theta_1}
= 1+ (-1)^{q+1} e^{i q \theta_1},
\label{A_expand}
\ee
which is derived from $f(\theta_1 + 2\pi/q,\theta_2)=f(\theta_1,\theta_2)$.

When $t$ satisfies
\be
0< t\le 2^{1/q},
\label{zero_cond}
\ee
the range of $f(\theta_1,\theta_2)$ is $0\le f (\theta_1,\theta_2) \le (t^q+2)^2$
and there exist two independent $(\theta_1,\theta_2)$'s with
$f(\theta_1,\theta_2)=0$.
From the secular equation (\ref{secular_equation}), we have two $E=0$
modes at these $(\theta_1,\theta_2)$'s. 
On the other hand, if $t$ satisfies
\be
t> 2^{1/q},
\ee
we have $(t^q-2)^2\le  f(\theta_1,\theta_2) \le (t^q+2)^2$ and there is
no $(\theta_1,\theta_2)$ with $f(\theta_1,\theta_2)=0$.
We have a gap around zero energy in this case.

Here we note that  the condition for the existence of zero modes for $\Phi=0$,
that is, $0<t\le 2$, is reproduced by (\ref{zero_cond}) with $q=1$.
(When $q=1$, (\ref{magnetic2}) reduces to that with $\Phi=0$.)

\subsection{Derivation from the normalizability condition of states}
\label{sec:states}

In Sec. \ref{sec:zero_mode}, we derived the condition for the existence 
of zero modes from the secular equation.
Here, we re-derive it from the normalizability condition of states 
with zero energy.

Let us first consider (\ref{magnetic4}) with $E=0$,
\be
\phi_n=-\frac{1}{t}A_n \phi_{n+1}, \quad
\psi_{n+1}=-\frac{1}{t}A_n^* \psi_{n}. 
\label{magnetic_zero3}
\ee
Then for $\Phi=p/q$, (\ref{A_expand}) and (\ref{magnetic_zero3}) lead to
\be
\phi_{Nq+l}=\left(-\frac{1}{t}\right)^q 
\left(\prod_{n=1}^{q}A_n \right)\phi_{(N+1)q+l}
=\left(-\frac{1}{t}\right)^q [1+(-1)^{q+1} e^{i q {\theta_1}}]
 \phi_{(N+1)q+l},
\nonumber\\
\psi_{(N+1)q+l}
=\left(-\frac{1}{t}\right)^q 
\left(\prod_{n=1}^{q}A_n^* \right)\psi_{Nq+l}
=\left(-\frac{1}{t}\right)^q 
[1+(-1)^{q+1} e^{-i q {\theta_1}}] \psi_{Nq+l},
\label{magnetic_zero5}
\ee
where $l=0,1,2,\ldots,q-1$.
Taking the absolute values of the both sides in (\ref{magnetic_zero5}),
we obtain
\be
|\phi_{Nq+l}|=\frac{2}{t^q}\left|\cos\left(\frac{q\theta_1}{2}+\frac{q+1}{2}\pi\right)\right| |\phi_{(N+1)q+l}|,
\nonumber\\
|\psi_{(N+1)q+l}|=\frac{2}{t^q}\left|\cos\left(\frac{q\theta_1}{2}+\frac{q+1}{2}\pi\right)\right| |\psi_{Nq+l}|.
\label{zero_absolute}
\ee
For $t^q > 2$, (\ref{zero_absolute}) gives
\be
|\phi_{Nq+l}| 
< |\phi_{(N+1)q+l}|, \quad
|\psi_{(N+1)q+l}| 
< |\psi_{Nq+l}|.
\label{diverge}
\ee
From (\ref{diverge}), we see that $|\phi_n|$ diverges for $n\to \infty$, and 
$|\psi_n|$ diverges for $n\to -\infty$.
Thus these states are not normalizable, and
no relevant zero modes exist.
We have a gap around $E=0$ in this case.
On the other hand, for $t^q\le 2$, (\ref{zero_absolute}) gives
\be
|\phi_{Nq+l}|=|\phi_{(N+1)q+l}|,\quad |\psi_{(N+1)q+l}|=|\psi_{Nq+l}|,
\ee
at 
$\theta_1= \pm \frac{2}{q}\cos^{-1}\frac{t^q}{2}+\frac{q+1}{q}\pi$.
Thus there exist two zero modes for $t\le 2^{1/q}$.
These results coincide with those of Sec. \ref{sec:zero_mode}.

\section{Spectrum near zero energy in a weak magnetic field}
\label{sec:small_magnetic}

In this section, we examine the spectrum near zero energy in a weak
magnetic field.
Although some numerical study was presented in Ref.\cite{MK06},
we perform detailed analytical study here.
On the basis of the condition for the existence of zero modes
obtained in the previous section,
we consider the following four cases separately:
\begin{enumerate}
\item[A.]
 $0<t\le 1$, where the condition (\ref{zero_cond})
is always satisfied and we have zero modes 
for all rational values of $\Phi$.
\item[B.]
$1 < t < 2$, where zero modes disappear and a gap around $E=0$ opens for $t>2^{1/q}$. 
\item[C.]
$t=2$, where one zero mode exists for $\Phi=0$.
\item[D.]
$t> 2$, where no zero modes exist.
\end{enumerate}

\subsection{$0<t\le 1$}
\label{sec:0<t<1}

\begin{figure}[t]
 \begin{center}
  \includegraphics[width=8.5cm,clip]{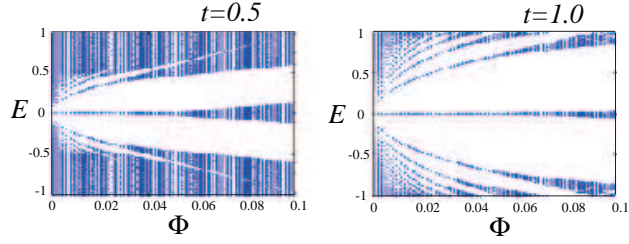}
\caption{\label{fig:t<=1}(Color online)
Energy bands as a function of $\Phi$ for $t=0.5$ and $t=1.0$.
}
\end{center}
\end{figure} 

\begin{figure}
 \begin{center}
  \includegraphics[width=8.5cm,clip]{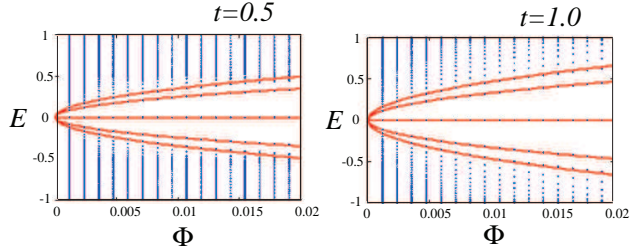}
\caption{\label{fig:t<=1_1}(Color online)
A closer look of Fig.\ref{fig:t<=1} in a weak magnetic field region.
 The energy levels (\ref{level_4}) and (\ref{level_3}) 
[or (\ref{level_2}) and (\ref{level_1})] are also shown by the red lines.
}
\end{center}
\end{figure} 

We show two examples of the energy bands 
as a function of $\Phi$ in Fig.\ref{fig:t<=1}. 
For $0<t\le 1$, we have zero modes for all rational values of $\Phi$.
As shown in the following, the energy bands in a weak magnetic 
field are well described by the continuum approximation.

In the continuum approximation, we use the Landau gauge
${\bf A}=(0,B x,0)$ for a magnetic field $B$. 
Then, substitution $p_x\to \hat{p}_x$ and $p_y\to \hat{p}_y+B x$ for (\ref{delta_1})
with $\hat{p}_x=-i\partial_x$, $\hat{p}_y=-i\partial_y$ and $B=\pi \Phi$
(see Appendix \ref{sec:appendixA}) 
gives the equation in a weak magnetic field as
\be
{\cal Q}_{\pm} \left(
\begin{array}{c} 
  \psi(x,y)\\ 
  \phi(x,y)
\end{array} \right)
=E
\left(
\begin{array}{c} 
  {\psi(x,y)}\\ 
  {\phi(x,y)}
\end{array} \right),
\label{Q_equation_pre}
\ee
where ${\cal Q}_{\pm}$ is given by
\be
{\cal Q}_{\pm}=
\left(
\begin{array} {cc}
0 &  {\cal D}_{\pm} \\
 {\cal D}_{\pm}^* & 0
\end{array} 
\right),
\quad
{\cal D}_{\pm}=-i t \hat{p}_x\pm\sqrt{4-t^2}(\hat{p}_y+\pi\Phi x).
\label{Hamiltonian_Q_pre}
\ee
Here, ${\cal D}_{+}$ and ${\cal D}_{-}$ are those
near ${\bm k}_{0}^{+}$ and ${\bm k}_{0}^{-}$, respectively.
Since ${\cal Q}_{\pm}$ and $\hat{p}_y$ commute each other,
we can replace $\hat{p}_y$ with a c-number $p_y$. 
Then putting $x\rightarrow x-p_y/\pi\Phi$, we obtain
\be
{\cal Q}_{\pm} \left(
\begin{array}{c} 
  \psi(x)\\ 
  \phi(x)
\end{array} \right)
=E
\left(
\begin{array}{c} 
  {\psi(x)}\\ 
  {\phi(x)}
\end{array} \right),
\label{Q_equation}
\ee
where 
\be
{\cal Q}_{\pm}=
\left(
\begin{array} {cc}
0 &  {\cal D}_{\pm} \\
 {\cal D}_{\pm}^* & 0
\end{array} 
\right),
\quad
{\cal D}_{\pm}=-i t \hat{p}_x \pm \sqrt{4-t^2} \pi \Phi x.
\label{Hamiltonian_Q}
\ee
From (\ref{Q_equation}) and (\ref{Hamiltonian_Q}), the following equation 
is obtained: 
\be
{\cal H}_{\pm}\left(
\begin{array}{c} 
  {\psi(x)}\\ 
  {\phi(x)}
\end{array} \right)
=E^2
 \left(
\begin{array}{c} 
  {\psi(x)}\\ 
  {\phi(x)}
\end{array} \right),
\label{H_0<t<2}
\ee
where
\be
{\cal H}_{\pm}={\cal Q}_{\pm}^2
=\left(
\begin{array} {cc}
 {\cal D}_{\pm} {\cal D}_{\pm}^* &  0 \\
 0 &  {\cal D}_{\pm}^* {\cal D}_{\pm}
\end{array} 
\right).
\label{Hamiltonian_1}
\ee
Therefore, we have 
\be
\left[t^2 \hat{p}_x^2+(4-t^2)\pi^2 \Phi^2 {x}^2 
- t \sqrt{4-t^2}\pi\Phi\sigma_z \right]
 \left(
\begin{array}{c} 
  {\psi(x)}\\ 
  {\phi(x)}
\end{array} \right)
=E^2 
\left(
\begin{array}{c} 
  {\psi(x)}\\ 
  {\phi(x)}
\end{array} \right), 
\label{H^2_K}
\ee
around ${\bm k}_{0}^{+}$, and 
\be
\left[t^2 \hat{p}_x^2+(4-t^2)\pi^2 \Phi^2 {x}^2 
+ t \sqrt{4-t^2}\pi\Phi\sigma_z \right]
\left(
\begin{array}{c} 
  {\psi(x)}\\ 
  {\phi(x)}
\end{array} \right)
=E^2
\left(
\begin{array}{c} 
  {\psi(x)}\\ 
  {\phi(x)}
\end{array} \right),
\label{H^2_K'}
\ee
around ${\bm k}_{0}^{-}$, where $\sigma_z$ is the $z$-component 
of the Pauli matrix.
\begin{figure}
 \begin{center}
  \includegraphics[width=8.5cm,clip]{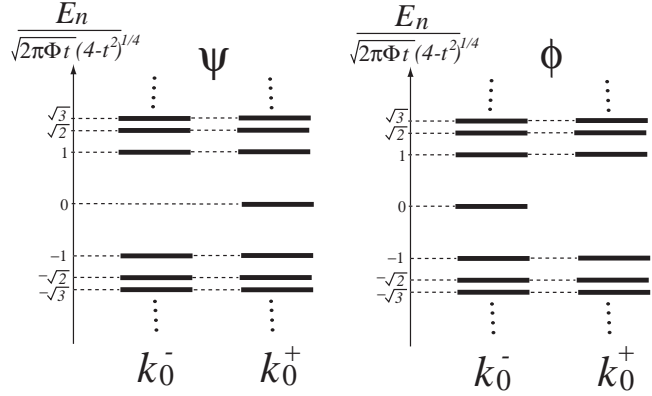}
\caption{\label{fig:energy_levels}
Energy levels around ${\bm k}_0^{+}$ and  ${\bm k}_0^{-}$.
}
\end{center}
\end{figure} 
Since the equations for $\psi$ in (\ref{H^2_K}) and (\ref{H^2_K'})
essentially coincide with those for harmonic oscillators, 
the energy level for ${\psi}$ around ${\bm k}_{0}^{+}$ is given by
\be
E_n =\pm\sqrt{2\pi\Phi t} (4-t^2)^{1/4} \sqrt{n},
\quad (n=0,1,2,\ldots),
\label{level_4}
\ee
and that around ${\bm k}_{0}^{-}$ is given by 
\be
E_n =\pm\sqrt{2\pi\Phi t} (4-t^2)^{1/4} \sqrt{n+1},
\quad (n=0,1,2,\ldots).
\label{level_3}
\ee
In a similar manner, the energy levels for $\phi$ 
around ${\bm k}_{0}^{+}$ and ${\bm k}_{0}^{-}$ are given by
\be
E_n =\pm\sqrt{2\pi\Phi t} (4-t^2)^{1/4} \sqrt{n+1},
\quad(n=0,1,2,\ldots),
\label{level_2}
\ee
and
\be
E_n =\pm\sqrt{2\pi\Phi t} (4-t^2)^{1/4} \sqrt{n},
\quad(n=0,1,2,\ldots),
\label{level_1}
\ee
respectively.
We show the energy levels around ${\bm k}_{0}^{+}$ and ${\bm k}_{0}^{-}$ 
in Fig.\ref{fig:energy_levels}.
As illustrated in Fig.\ref{fig:t<=1_1}, 
the energy bands for $0<t\le 1$ 
come to be well fitted by (\ref{level_4}) and (\ref{level_3}) 
[or (\ref{level_2}) and (\ref{level_1})]
in a weak magnetic field.

\subsection{$1< t < 2$}

\begin{figure}
 \begin{center}
  \includegraphics[width=7.0cm,clip]{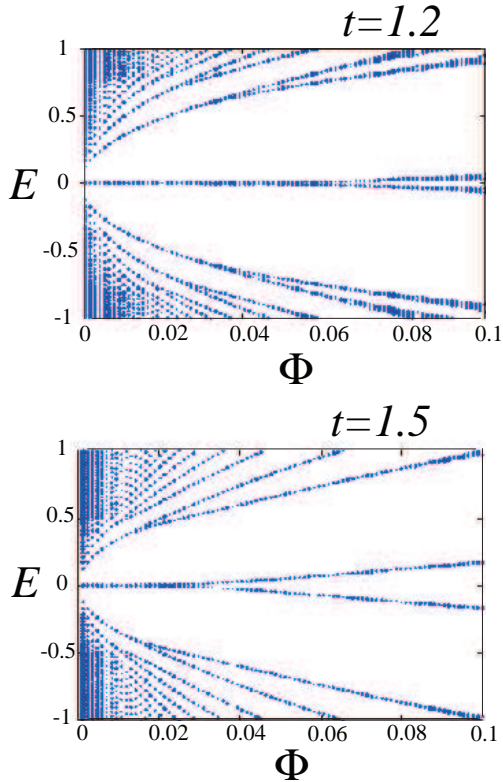}
\caption{\label{fig:1<t<2_1}(Color online)
Energy bands as a function of $\Phi$ for $t=1.2$ and $t=1.5$.
}
\end{center}
\end{figure} 

For $1<t<2$, a gap around $E=0$ opens for $t>2^{1/q}$.
This implies that {\it in a weak magnetic field} ($q \gg 1$),
{\it a gap around $E=0$ opens by a tiny distortion of graphene,
$t=1+\epsilon$ $(0<\epsilon\ll 1)$}.
Since we do not have $E=0$ states,
the expressions (\ref{level_4}) and (\ref{level_1}) need to be modified.
We show two examples of energy bands 
as a function of $\Phi$ for $1<t<2$ in Fig.\ref{fig:1<t<2_1}.

\begin{figure}
 \begin{center}
  \includegraphics[width=7.0cm,clip]{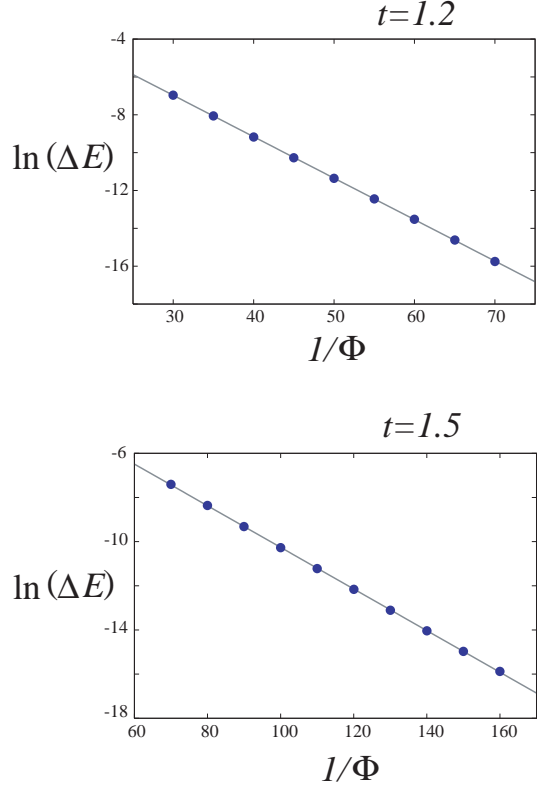}
\caption{\label{fig:1<t<2_2}(Color online)
The natural logarithm of the gap around $E=0$ 
as a function of $1/\Phi$ for $t=1.2$ and $t=1.5$.
The slope of the fitting line gives $-\alpha$.}
\end{center}
\end{figure} 

Let us focus on the gap around $E=0$.
In order to see how it behaves,
we plot the natural logarithm of the gap $\Delta E$ around $E=0$
as a function of $1/\Phi$ in Fig.\ref{fig:1<t<2_2}.
From this, we find that it behaves as
\be
\Delta E \sim \exp(-\alpha/\Phi).
\label{exp_gap}
\ee
The values of $\alpha$ are obtained from Fig.\ref{fig:1<t<2_2}
as $\alpha\sim 0.22$ and $0.094$ 
for $t=1.2$ and $1.5$, respectively.

The non-analytic behavior (\ref{exp_gap}) can be understood
as breaking of supersymmetry \cite{Witten81,Witten81_2,MS99}
in our model.
The operator ${\cal Q}_{\pm}$ transforms $\psi$ to $\phi$ and vice versa,
which is seen from (\ref{Q_equation}) and (\ref{Hamiltonian_Q}).
By identifying ${\cal Q}_{\pm}$ with generators of supersymmetry, ${\cal
H}_{\pm}$ in (\ref{H_0<t<2}) can be considered as sypersymmetric Hamiltonians,
${\cal H}_{\pm}={\cal Q}_{\pm}^2$ ($\phi$ is ``boson'', 
and $\psi$ is ``fermion'').
Due to the supersymmetry, there is no
perturbative (or power-law) correction with respect to $\Phi$
for the $E=0$ states.
However, tunneling effects break the supersymmetry spontaneously and 
the non-perturbative correction (\ref{exp_gap}) appears as a gap around $E=0$.

When two Dirac zero modes at $\Phi=0$ are close to each other 
in the momentum space, namely, $t\sim 2$,
the gap around $E=0$ can be estimated by the WKB method.
For $t\sim 2$, the two Dirac zero modes at $\Phi=0$ 
are located at (\ref{zero_modes}) with
\be
\cos^{-1}\frac{t}{2} \sim \sqrt{2-t}\equiv G,
\ee
and for $t=2$, they merge into a confluent point (\ref{confluent}).
For $t\sim 2$, it is convenient to expand $k_x$ and $k_y$ around the
confluent point (\ref{confluent}) instead of ${\bm k}_{0}^{+}$ or ${\bm k}_{0}^{-}$:
\be
k_x=k_x^*+p_x=\pi+p_x, \quad
k_y=k_y^*+p_y=p_y,
\quad (|p_x|,|p_y|\ll 1).
\ee
Then ${\cal D}({\bm k})$ in (\ref{delta_zero}) is given by
\be
{\cal D}({\bm k})=-2ip_x+p_y^2-G^2.
\label{delta_near}
\ee

In a weak $\Phi$,
we can use the continuum approximation.
We use the Landau gauge ${\bf A}=(0,B x,0)$ for a magnetic field $B$. 
Then, substitution $p_x\to\hat{p}_x$ and $p_y\to \hat{p}_y+B x$ for (\ref{delta_near})
with $\hat{p}_x=-i\partial_x$, $\hat{p}_y=-i\partial_y$ and  $B=\pi \Phi$ (see Appendix \ref{sec:appendixA})
gives the following equation,
\be
{\cal Q} \left(
\begin{array}{c} 
  \psi(x,y)\\ 
  \phi(x,y)
\end{array} \right)
=E
\left(
\begin{array}{c} 
  {\psi(x,y)}\\ 
  {\phi(x,y)}
\end{array} \right),
\label{Q_equation_t_2_pre}
\ee
where ${\cal Q}$ is given by
\be
{\cal Q}=
\left(
\begin{array} {cc}
0 &  {\cal D} \\
 {\cal D}^* & 0
\end{array} 
\right),
\quad
{\cal D}=-2 i \hat{p}_x +( \hat{p}_y+\pi\Phi x)^2-G^2.
\label{Hamiltonian_Q_t_2_pre}
\ee
In a similar manner as Sec.\ref{sec:0<t<1}, we replace $\hat{p}_y$ with a
c-number $p_y$ and put $x\to x-p_y/\pi\Phi$. 
Then we obtain
\be
{\cal Q} \left(
\begin{array}{c} 
  {\psi(x)}\\ 
  {\phi(x)}
\end{array} \right)
=E
\left(
\begin{array}{c} 
  {\psi(x)}\\ 
  {\phi(x)}
\end{array} \right),
\label{Q_equation_t_2}
\ee
where ${\cal Q}$ is given by
\be
{\cal Q}=
\left(
\begin{array} {cc}
0 &  {\cal D} \\
 {\cal D}^* & 0
\end{array} 
\right),
\quad
{\cal D}=-2i \hat{p}_x +\pi^2\Phi^2 x^2 - G^2.
\label{Hamiltonian}
\ee
\begin{figure}
 \begin{center}
  \includegraphics[width=8.5cm,clip]{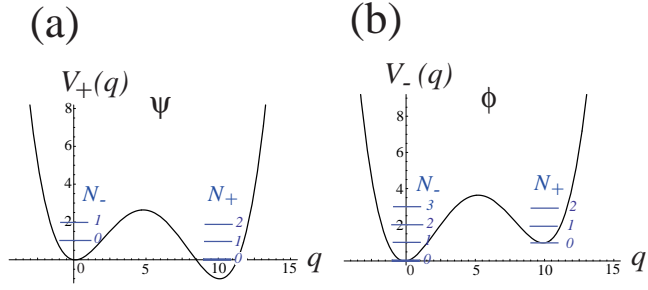}
\caption{\label{fig:potential_v}(Color online)
The asymmetric double-well potential
(a) given by (\ref{potential_1}) with $g=0.1$, and
(b) given by (\ref{potential_2}) with $g=0.1$.
Energy levels around $q=0$ and $q=1/g$ are also shown.}
\end{center}
\end{figure} 
Identifying ${\cal Q}$ with a generator of supersymmetry, we have the
supersymmetric Hamiltonian ${\cal H}={\cal Q}^2$, which satisfies
\be
{\cal H}\left(
\begin{array}{c} 
  {\psi(x)}\\ 
  {\phi(x)}
\end{array} \right)
=E^2 \left(
\begin{array}{c} 
  {\psi(x)}\\ 
  {\phi(x)}
\end{array} \right).
\label{H_near2}
\ee
By using the following variable $q$, 
\begin{eqnarray}
x+\frac{G}{\pi\Phi}=q\sqrt{\frac{1}{\pi G\Phi}},
\end{eqnarray}
(\ref{H_near2}) can be rewritten as
\be
\left[-\frac{1}{2}\frac{d^2}{d q^2}
+\frac{1}{2}q^2(1-gq)^2
-\left(gq-\frac{1}{2}\right)\sigma_z \right]
\left(
\begin{array}{c} 
  {\psi(q)}\\ 
  {\phi(q)}
\end{array} \right)
={\cal E} 
\left(
\begin{array}{c} 
  {\psi(q)}\\ 
  {\phi(q)}
\end{array} \right),
\label{susy_1}
\ee
with
\be
g=\frac{1}{2G}\sqrt{\frac{\pi\Phi}{G}},
\quad
{\cal E}=\frac{1}{8 \pi\Phi G}E^2.
\label{g}
\ee
Therefore, the potential terms for ${\psi(x)}$ and ${\phi(x)}$
are given by
\be
V_+(q)=\frac{1}{2}q^2(1-gq)^2-gq,
\label{potential_1}
\ee
and 
\be
V_-(q)=\frac{1}{2}q^2(1-gq)^2+gq,
\label{potential_2}
\ee
respectively (see Fig.\ref{fig:potential_v}).

In the leading order of $g$, 
the potentials (\ref{potential_1}) and (\ref{potential_2})
are well approximated by the harmonic oscillator
 around $q=0$ and $q=1/g$.
Around $q=0$, (\ref{susy_1}) becomes
\be
\left[-\frac{1}{2}\frac{d^2}{d q^2}
+\frac{1}{2}q^2+\frac{1}{2}\sigma_z \right] 
\left(
\begin{array}{c} 
  {\psi(q)}\\ 
  {\phi(q)}
\end{array} \right)
={\cal E} 
\left(
\begin{array}{c} 
  {\psi(q)}\\ 
  {\phi(q)}
\end{array} \right),
\label{susy_q=0}
\ee
and around $q=1/g$, (\ref{susy_1}) becomes
\be
\left[-\frac{1}{2}\frac{d^2}{d q^2}
+\frac{1}{2}\left(q-\frac{1}{g}\right)^2
-\frac{1}{2}\sigma_z \right] 
\left(
\begin{array}{c} 
  {\psi(q)}\\ 
  {\phi(q)}
\end{array} \right)
={\cal E}
\left(
\begin{array}{c} 
  {\psi(q)}\\ 
  {\phi(q)}
\end{array} \right).
\label{susy_q=1/g}
\ee
Therefore, the energy levels for ${\psi}$ around $q=1/g$ and $q=0$
are given by
\be
{\cal E}_{N_+}=N_{+},
\quad (N_{+}=0,1,2,\ldots),
\label{psi_level1}
\ee
and 
\be
{\cal E}_{N_-}=N_{-}+1,
\quad (N_{-}=0,1,2,\ldots),
\label{psi_level2}
\ee
respectively,
and those for ${\phi}$ around $q=1/g$ and $q=0$
are given by
\be
{\cal E}_{N_+}=N_{+}+1,
\quad (N_{+}=0,1,2,\ldots),
\label{phi_level1}
\ee
and 
\be
{\cal E}_{N_-}=N_{-},
\quad (N_{-}=0,1,2,\ldots),
\label{phi_level1}
\ee
respectively.

Now take into account tunneling effects between the energy levels 
around $q=0$ and $q=1/g$.
The tunneling effects can be estimated by the WKB method 
presented in Appendix D of Ref.\cite{MS99}.
Here we consider only the equation for ${\psi(q)}$ since 
the equation for ${\phi(q)}$ gives the same result.
The solution of (\ref{susy_q=0}) which vanishes for $q\to -\infty$ is
given by
\be
{\psi}(q)=A D_{\nu}\left(-\sqrt{2} q \right),
\label{WKB_solution1}
\ee
where $\nu={\cal E}-1$, $A$ is a constant,
and $D_{\nu}$ the parabolic cylinder function\cite{parabolic}.
The solution of (\ref{susy_q=1/g}) which vanishes for $q\to \infty$ is
\be
{\psi}(q)=\tilde{A} D_{\nu+1}\left[\sqrt{2} \left(q-1/g \right)\right],
\label{WKB_solution2}
\ee
where $\tilde{A}$ is a constant.
We connect these solutions (\ref{WKB_solution1}) and (\ref{WKB_solution2})
with that in the forbidden region.
In the forbidden region, 
the usual semi-classical expression for the wave function
is available:
\be
{\psi}(q)=
\frac{C_1}{\sqrt{k(q)}}\exp\left(-\int_{q_1}^q k(x)dx\right)
+\frac{C_2}{\sqrt{k(q)}}\exp\left(\int_{q_1}^q k(x)dx\right),
\label{WKB_forbidden}
\ee
where $k(q)=\sqrt{2(V_+(q)-{\cal E})}$ with $V_+(q)$ in (\ref{potential_1}),
$q_i$ $(i=1,2)$ are the turning points, $V_+(q_i)={\cal E}$, and 
$C_i$ $(i=1,2)$ are constants.
Connecting (\ref{WKB_solution1}) with (\ref{WKB_forbidden}),
and (\ref{WKB_solution2}) with (\ref{WKB_forbidden}),
we obtain
\be
\gamma^2\left(-\frac{2}{g^2}\right)^{2{\cal E}-1} 
\Gamma\left(1-{\cal E} \right)
\Gamma\left(-{\cal E}\right)=1,
\quad
\gamma=\frac{e^{-1/6g^2}}{g\pi^{1/2}}.
\label{gamma}
\ee
For ${\cal E}$ near zero energy (${\cal E}\ll 1$), we have
\be
\left(-\frac{2}{g^2}\right)^{2{\cal E}-1}\simeq -\frac{g^2}{2},
\quad
\Gamma\left(1-{\cal E}\right)\simeq 1,
\quad
\Gamma\left(-{\cal E}\right)\simeq -\frac{1}{{\cal E}},
\label{E_approximation}
\ee
thus the solution of (\ref{gamma}) for $g\ll 1$ is obtained as
\be
{\cal E}\left(=\frac{1}{8 \pi\Phi G}{E}^2\right)=\gamma^2\frac{g^2}{2}.
\label{WKB}
\ee
This implies that the gap around $E=0$ is given by
\be
\Delta E= 4\sqrt{\Phi G}
\exp\left(-\frac{2 G^3}{3 \pi \Phi}\right),
\label{WKB_result}
\ee
and the exponent $\alpha$ in (\ref{exp_gap}) is given by
\be
\alpha=\frac{2}{3\pi}G^3=\frac{2}{3\pi}(2-t)^{3/2}\equiv \alpha_{\rm WKB},
\quad
(0<2-t\ll 1).
\label{alpha_WKB1}
\ee

\begin{figure}
 \begin{center}
  \includegraphics[width=7.0cm,clip]{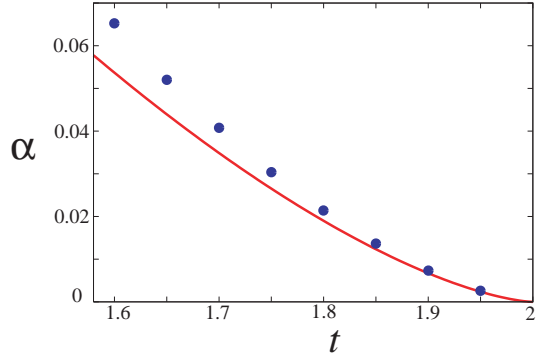}
\caption{\label{fig:alpha}(Color online)
$\alpha$ as a function of $t$,
where those obtained from numerical calculations
and the WKB analysis are shown
 by points and the line, respectively.}
\end{center}
\end{figure} 

\begin{table}
\begin{center}
\caption{$\alpha$ for several $t$
 obtained from numerical calculations and the WKB analysis.
The relative discrepancies between them, $\delta$, are also shown.
}
\label{alpha_delta}
\begin{tabular}{|c||c|c|c|}
\multicolumn{4}{c}{}\\
\hline
$t$ &  $\alpha$ & $\alpha_{\rm WKB}$ & $\delta$ \\
\hline\hline
1.6  & $6.523\times 10^{-2}$ & $5.368\times 10^{-2}$ & 0.177 \\
\hline
1.65 & $5.198\times 10^{-2}$ & $4.394\times 10^{-2}$ & 0.155 \\
\hline
1.7 & $4.076\times 10^{-2}$ & $3.487\times 10^{-2}$ & 0.145  \\
\hline
1.75 & $3.042\times 10^{-2}$ & $2.653\times 10^{-2}$ & 0.128 \\
\hline
1.8 & $2.140\times 10^{-2}$ & $1.898\times 10^{-2}$ & 0.113\\
\hline
1.85 & $1.369\times 10^{-2}$ & $1.233\times 10^{-2}$ & 0.0993 \\
\hline
1.9 & $7.294\times 10^{-3}$ & $6.711\times 10^{-3}$ & 0.0799\\
\hline
1.95 & $2.555\times 10^{-3}$ & $2.373\times 10^{-3}$ & 0.0712\\
\hline
\end{tabular}
\end{center}
\end{table}

Now we compare (\ref{alpha_WKB1}) with those obtained numerically
for a small $\Phi$.
In Fig.\ref{fig:alpha}, we show $\alpha$ as a function of $t$,
and in Table \ref{alpha_delta}, we list them.
The relative discrepancy 
\be
\delta 
\equiv \left|\frac{\alpha_{\rm WKB}-\alpha}{\alpha}\right|,
\ee 
decreases as $t$ approaches $2$.
This is because the neglected terms $O(p_y^4)$
in (\ref{delta_near}) come to be smaller and smaller as $t$ approaches $2$.

\subsection{$t=2$}
\begin{figure}
 \begin{center}
 \includegraphics[width=8.5cm,clip]{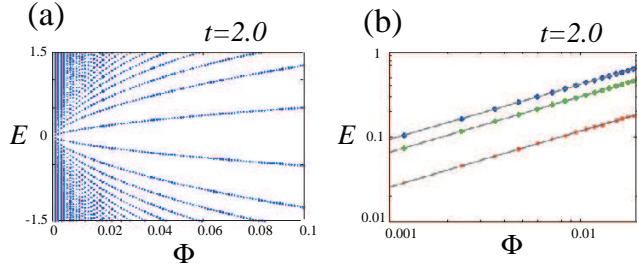}
\caption{\label{fig:t=2.0_1}(Color online)
(a)Energy bands as a function of $\Phi$ for $t=2.0$.
(b)The three lowest bands in $E\ge 0$ in the log-log scale.
}
\end{center}
\end{figure} 

At $t=2$, the two Dirac zero modes at $\Phi=0$ 
merge into the confluent point (\ref{confluent}).
As a consequence, a gap around $E=0$ in a weak magnetic field 
makes a transition from an exponential (non-perturbative)
to a power-law (perturbative) behavior as a function of $\Phi$.

In Fig.\ref{fig:t=2.0_1}(a), we show the energy bands
as a function of $\Phi$ for $t=2.0$.
We show the three lowest bands in $E\ge 0$
in the log-log scale for weak magnetic field in Fig.\ref{fig:t=2.0_1}(b).
We fit our data by
\be
E\sim \Phi^{\kappa}.
\ee
For $t=2.0$, we obtain the exponent $\kappa$ as 
$\kappa\sim 0.66$, $0.65$, and $0.65$ for 
the lowest, the second lowest, and the third lowest bands in $E\ge 0$,
respectively.
Thus for $t=2.0$ we have a power-law behavior $E \sim \pm \Phi^{2/3}$.

The behavior $E \sim \pm \Phi^{2/3}$ 
is derived analytically from a particular dispersion relation 
at $t=2$ for $\Phi=0$ \cite{PD08}.
For $\Phi=0$, (\ref{eigen_energy}) and (\ref{delta_near}) with $G=0$ give
\be
E=\pm \sqrt{4 p_x^2 +  p_y^4},
\label{dispersion_t=2}
\ee
which is linear in one direction and quadratic in the other.
The exponent $\kappa$ is obtained from $S(E)\sim \Phi$,
where $S(E)$ is the area surrounded by an orbit of energy $E$
in the momentum space \cite{PD08}.
From (\ref{dispersion_t=2}), we have
\be
S(E)=\frac{\Gamma(1/4)^2}{3\sqrt{2\pi}} |E|^{3/2}, 
\ee
thus $E \sim \pm \Phi^{2/3}$.

\subsection{$t>2$}
\begin{figure}
 \begin{center}
  \includegraphics[width=8.5cm,clip]{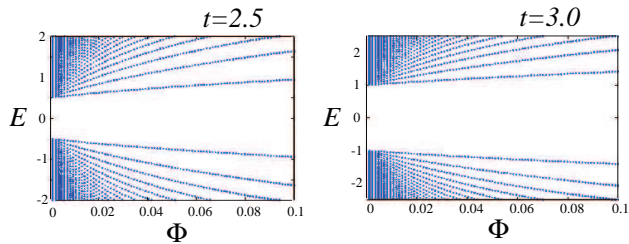}
\caption{\label{fig:t>2_1}(Color online)
Energy bands as a function of $\Phi$
for $t=2.5$ and $t=3.0$.
}
\end{center}
\end{figure} 
\begin{figure}
 \begin{center}
  \includegraphics[width=8.5cm,clip]{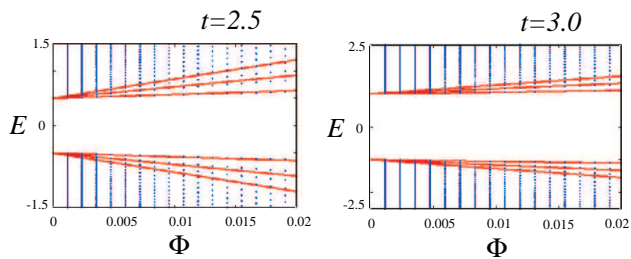}
\caption{\label{fig:t>2_linear}(Color online)
A closer look of Fig.\ref{fig:t>2_1} in a weak magnetic field region.
The expressions (\ref{levels_t>2}) for $n=0,1,2$ are also shown 
by the red lines.}
\end{center}
\end{figure} 
 
For $t>2$, we do not have zero modes 
but have a gap around $E=0$. 
In Fig.\ref{fig:t>2_1}, we show two examples of
the energy bands as a function of $\Phi$ for $t>2$.

Let us study behavior of energy bands in a weak magnetic field  
by the continuum approximation.
For $\Phi=0$, we expand $k_x$ and $k_y$ around $(k_x,k_y)=(\pi,0)$,
\be
k_x=\pi+p_x,\quad k_y=p_y, \quad (|p_x|,|p_y|\ll 1),
\ee
then (\ref{eigen_energy}) gives
\be
E=\pm \sqrt{E_g^2+2t p_x^2+2(t-2) p_y^2},
\quad E_g=t-2.
\label{energy_t>2}
\ee
Thus for small $|p_x|$ and $|p_y|$ ($|p_x|,|p_y|\ll E_g$),
(\ref{energy_t>2}) is written as
\be
E =\pm \left(E_g+\frac{t}{t-2}p_x^2+p_y^2\right),
\label{energy_t>2_2}
\ee
which is quadratic in both $p_x$ and $p_y$.
In the continuum approximation, we put $p_x\to \hat{p}_x$ and $p_y\to
\hat{p}_y+Bx$ with $\hat{p}_x=-i\partial_x$, $\hat{p}_y=-i\partial_y$
and $B=\pi\Phi$.
Then we have 
\be
E=\pm E_g \left[1+2\pi\sqrt{t}
\left(n+\frac{1}{2} \right)\frac{\Phi}{(t-2)^{3/2}}\right],
\quad (n=0,1,2,\ldots),
\label{levels_t>2}
\ee  
where we have neglected higher order corrections of
$O \left(\left(\frac{2\pi\sqrt{t}(n+1/2)\Phi}{(t-2)^{3/2}}\right)^2 \right)$.
The energy bands in the weak magnetic field limit
$\left(\Phi \ll \frac{(t-2)^{3/2}}{2\pi\sqrt{t} (n+1/2)}\right)$
are well approximated by (\ref{levels_t>2}), 
which is seen in Fig.\ref{fig:t>2_linear}.
We note that, for $t \sim 2$, the neglected higher order corrections of 
$O \left(\left(\frac{2\pi\sqrt{t}(n+1/2)\Phi}{(t-2)^{3/2}}\right)^2 \right)$
can not be neglected.
However, they become small for $t\gg 2$, and the energy bands 
in a weak magnetic field are well fitted by (\ref{levels_t>2}).
This result is consistent with the fact that the honeycomb lattice 
becomes equivalent to the square lattice for $t\gg 2$\cite{MK07,MS08}.

\begin{figure}
 \begin{center}
 \includegraphics[width=8.0cm,clip]{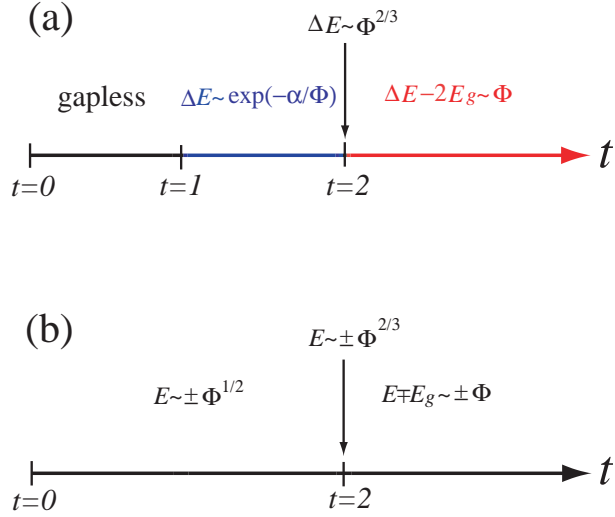}
\caption{\label{fig:diagram}(Color online)
(a)Behavior of a gap $\Delta E$ around $E=0$
as a function of $t$ in a weak magnetic field.
(The nearest bands to $E=0$ show the same behavior as $\Delta E$.)
(b)Behavior of the other energy bands for $E\approx 0$ 
as a function of $t$ in a weak magnetic field.
}
\end{center}
\end{figure}

We summarize our results of this section in Fig.\ref{fig:diagram}.

\section{$E=0$ edge states in anisotropic honeycomb lattice}
\label{sec:edge}
\begin{figure}
 \begin{center}
 \includegraphics[width=7.5cm,clip]{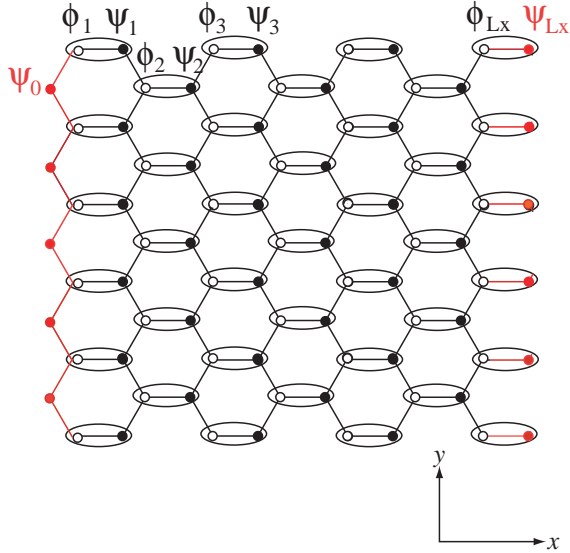}
\caption{\label{fig:honeycomb_edge}(Color online)
Honeycomb lattices with zigzag and bearded edges ($L_x=7$).
The left and the right edges are the zigzag and the bearded edges, 
respectively.}
\end{center}
\end{figure} 

In this section, we examine $E=0$ edge states.
The condition for the existence of zero energy edge states 
in a magnetic field is analytically derived.
On the basis of it, it turns out that the anisotropy of the hopping integrals
induces abrupt changes of the number of zero energy edge states,
which depend on the shapes of the edges sensitively.

In order to see this, we focus on the honeycomb lattices 
with zigzag and bearded edges as shown in Fig.\ref{fig:honeycomb_edge}.  
For these lattices, we have edges along the $y$-direction.  
Let us impose the periodic boundary condition along the $y$-direction:
\be
\psi_{n,m+2L_y}=\psi_{n,m},\quad \phi_{n,m+2L_y}=\phi_{n,m},
\ee
where an integer $L_y$ denotes the circumference of the cylinder.
Then one can write
\be
\psi_{n,m}=\exp\left(i\frac{k_y}{2} m\right)\psi_n,
\quad
\phi_{n,m}=\exp\left(i\frac{k_y}{2}m \right)\phi_{n}.
\label{edge_eq2}
\ee
Let us focus on $E=0$ states.
In the same manner as Sec.\ref{sec:states}, for $E=0$ and $\Phi=p/q$ 
with coprime integers $p$ and $q$,
the amplitudes of wavefunctions separated by a distance $q$ 
are found to satisfy
\be
|\psi_{(N+1)q+l}|=r |\psi_{Nq+l}|,\quad |\phi_{Nq+l}|=r |\phi_{(N+1)q+l}|,
\nonumber\\ (l=0,1,2,\ldots,q-1),
\label{N_zero}
\ee
with
\be
r\equiv\frac{2}{t^q}
\left|\cos\left(q\frac{k_y}{2}+\frac{q+1}{2}\pi\right)\right|.
\label{r}
\ee
From (\ref{N_zero}), we have
\be
|\psi_{Nq+l}|=r^N |\psi_{l}|,\quad |\phi_{l}|=r^N |\phi_{Nq+l}|.
\label{N_zero2}
\ee
The boundary conditions for the zigzag and the bearded edges are given by
\be
\phi_0=0,\quad \phi_{Lx+1}=0,
\label{eq:b}
\ee
respectively. These boundary conditions give $\phi_n=0$ for all $n$, 
thus we consider only $\psi_n$ in the following.

Suppose that $L_x$ is large enough: $L_x\gg q$.
Then from (\ref{N_zero2}), if $r<1$ ($r>1$) we have $E=0$ states 
on the zigzag edge (bearded edge). 
For $t>2^{1/q}$, $r<1$ is satisfied 
for all values of $k_y$, but $r>1$ is not satisfied for any values of $k_y$. 
Thus we have $E=0$ states localized on the zigzag edge
for all values of $k_y$, but we do not have $E=0$ states localized on the
bearded edge for any values of $k_y$.
For $t< 2^{1/q}$, $E=0$ edge states exist both on the zigzag 
and the bearded edge.
The total width $d_{\rm zigzag}(t,q)$ of the region of $k_y$ which gives 
$E=0$ states on the zigzag edge is given by
\be
d_{\rm zigzag}(t,q)
=2\left(\pi-2\cos^{-1}\frac{t^q}{2}\right),
\ee
and that for the bearded edge, $d_{\rm bearded}(t,q)$, is given by
\be
d_{\rm bearded}(t,q)=2\pi-d_{\rm zigzag}(t,q)
=4\cos^{-1}\frac{t^q}{2}.
\ee

\begin{figure}
 \begin{center}
 \includegraphics[width=7.0cm,clip]{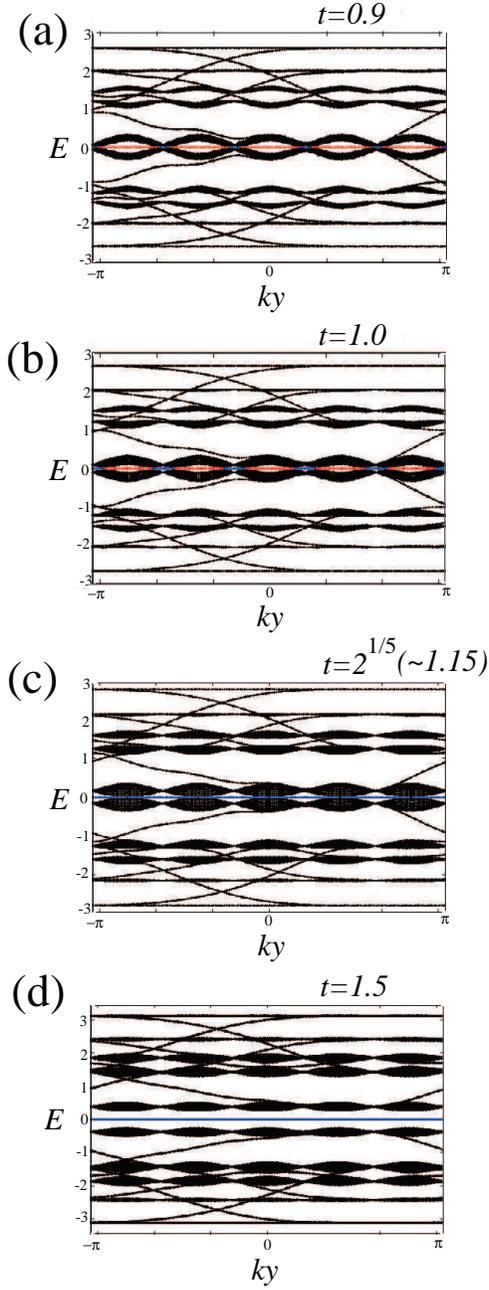}
\caption{\label{fig:zigzag_bearded_q=5}(Color online)
Energy spectra of honeycomb lattices with zigzag and bearded edges
for $\Phi=1/5$ and $L_x=50$.
(a) $t=0.9$, (b) $t=1.0$, (c) $t=2^{1/5}(\sim 1.15)$ (d) $t=1.5$.
The $E=0$ edge states localized on the zigzag edges are on the blue lines,
and those localized on the bearded edges are on the red lines.}
\end{center}
\end{figure}

\begin{figure}
 \begin{center}
 \includegraphics[width=7.0cm,clip]{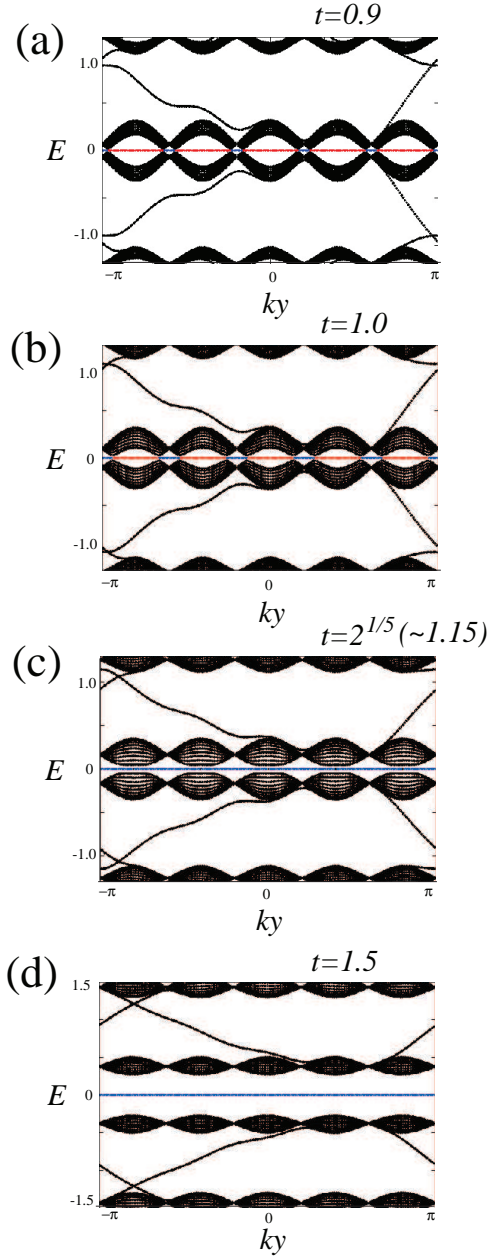}
\caption{\label{fig:zigzag_bearded_q=5_close}(Color online)
A closer look of Fig.\ref{fig:zigzag_bearded_q=5} at $E\approx 0$.}
\end{center}
\end{figure}

For a fixed value of $\Phi$,
$d_{\rm zigzag}$ ($d_{\rm bearded}$) 
increases (decreases) as $t$ increases in the region of $0<t<2^{1/q}$.
At $t=2^{1/q}$, $d_{\rm zigzag}$ covers the whole region of $k_y$, and
$d_{\rm bearded}$ vanishes.
For $t>2^{1/q}$,
we have $E=0$ edge states on the zigzag edges for all values of $k_y$. 
We show examples of the energy spectra of honeycomb lattices 
with zigzag and bearded edges in Figs.\ref{fig:zigzag_bearded_q=5} and
\ref{fig:zigzag_bearded_q=5_close}.

Effects of the anisotropy of the hopping integrals are evident
in a weak magnetic field, $\Phi\ll 1$ ($q\gg 1$).
For $t=1$, $d_{\rm zigzag}$ and $d_{\rm bearded}$ 
do not depend on the magnetic field,
and are given by $d_{\rm zigzag}=\frac{2\pi}{3}$ and 
$d_{\rm bearded}=\frac{4\pi}{3}$, respectively.
However, for $t<1$, $d_{\rm zigzag}$ decreases toward $0$ 
and $d_{\rm bearded}$ increases toward $2\pi$
as $q$ increases.
In contrast, for $t>1$, $d_{\rm zigzag}$ ($d_{\rm bearded}$) 
increases (decreases) as $q$ increases
and reaches $2\pi$ ($0$) at $q=\ln 2/\ln t$.
Note that even in a small anisotropy, $d_{\rm zigzag}$ and $d_{\rm
bearded}$ change abruptly in a weak magnetic field ($q\gg 1$).

Instead of $d_{\rm zigzag}$ and $d_{\rm bearded}$,
we also consider the integrated charge density $I_n$ for $E=0$
edge states\cite{MA08,MA08_2}:
\be
I_n=\int  {|\psi_n(k_y)|^2 dk_y},
\ee
where the normalization condition is imposed on $\psi_n(k_y)$,
\be
\sum_{n=0}^{L_x} |\psi_n(k_y)|^2=1.
\label{normalize}
\ee
To characterize the numbers of the states localized on the edges, we
introduce
the following quantities,
\be
N_{\rm zigzag}=\sum_{n=0}^{q-1} I_n,\quad 
N_{\rm bearded}=\sum_{n=L_x-q+1}^{L_x} I_n.
\label{N_zig_bea}
\ee
For $r<1$, only the zigzag edge has $E=0$ modes, and
$N_{\rm zigzag}$ is evaluated as
\be
N_{\rm zigzag}
=\int\left(1-\frac{2}{t^{2q}}\right) d k_y 
+(-1)^{q}\frac{2}{t^{2q}}\int \cos(q k_y) d k_y,
\label{Nz2}
\ee
where the domain of the integration is restricted to those $k_y$ with $r<1$.
Here we have used the relation derived from 
(\ref{N_zero2}) and (\ref{normalize}):
\be
\sum_{j=0}^{ \lfloor{(L_x+1)/q}\rfloor -1} r^{2j} \sum_{n=0}^{q-1} |\psi_n(k_y)|^2
+O(r^{2 L_x/q})=1,
\label{I2_pre}
\ee
where $O(r^{2 L_x/q})$ corrections can be neglected since $r<1$ and $L_x\gg q$.
($\lfloor{x}\rfloor$ denotes the integer part of $x$.)
Eq. (\ref{I2_pre}) is rewritten as
\be
\sum_{n=0}^{q-1} |\psi_n(k_y)|^2=1-r^2+O(r^{2 L_x/q}),
\label{I2}
\ee
then substituting this into the first equation in (\ref{N_zig_bea})
and using (\ref{r}), we obtain Eq. (\ref{Nz2}).
On the other hand, for $r>1$, we have $E=0$ modes only on the bearded
edge, and $N_{\rm bearded}$ is given by
\be
N_{\rm bearded}
=\int d k_y -\frac{t^{2q}}{4}\int \frac{1}{\cos^2
\left(q\frac{k_y}{2}+\frac{q+1}{2}\pi\right)} d k_y,
\label{Nb2}
\ee
where the domain of the integration is restricted to those $k_y$ with $r>1$.
Here we have used the relation
\be
\sum_{n=L_x-q+1}^{L_x} |\psi_n(k_y)|^2=1-r^{-2}+O(r^{-2 L_x/q}),
\label{I4}
\ee
which is derived from (\ref{N_zero2}) and (\ref{normalize})
in a similar manner as Eq. (\ref{I2}).

Let us now evaluate $N_{\rm zigzag}$ and $N_{\rm bearded}$ from
(\ref{Nz2}) and (\ref{Nb2}).
For $t>2^{1/q}$, $r<1$ is realized for all values
of $k_y$ as shown above.  
Therefore, 
\be
N_{\rm zigzag}
&=&\int_{0}^{2\pi}\left(1-\frac{2}{t^{2q}}\right) d k_y
+(-1)^{q}\frac{2}{t^{2q}}\int_{0}^{2\pi} \cos(q k_y) d k_y\nonumber\\ 
&=&2\pi\left(1-\frac{2}{t^{2q}}\right),
\nonumber\\
N_{\rm bearded}
&=&0.
\ee
For $t< 2^{1/q}$, either $r>1$ or $r<1$ is realized 
by a suitable choice of $k_y$.
Thus both $N_{\rm zigzag}$ and $N_{\rm bearded}$ become nonzero as, 
\be
N_{\rm zigzag}
&=& d_{{\rm zigzag}} \left(1-\frac{2}{t^{2q}}\right)- \frac{4}{t^{2q}}\int_{2\cos^{-1}\frac{t^q}{2}}^{\pi}  \cos k_y d k_y \nonumber\\
&=&2\left(\pi-2\cos^{-1}\frac{t^q}{2}\right)\left(1-\frac{2}{t^{2q}}\right)
+\frac{4}{t^q}\sin\left(\cos^{-1}\frac{t^q}{2}\right), 
\nonumber\\
N_{\rm bearded}
&=&d_{{\rm bearded}} -\frac{t^{2q}}{2}\int_{0}^{2\cos^{-1}\frac{t^q}{2}} \frac{1}{\cos^2 \left(\frac{k_y}{2}\right)} d k_y  \nonumber\\
&=&4\cos^{-1}\frac{t^q}{2}- 2 t^{q}\sin\left(\cos^{-1}\frac{t^q}{2}\right).
\ee
We also find that for $t=1$, $N_{\rm zigzag}$ and $N_{\rm bearded}$ are independent of $\Phi$,
\be
N_{\rm zigzag}=2\sqrt{3}-\frac{2\pi}{3},
\quad
N_{\rm bearded}=\frac{4}{3}\pi-\sqrt{3}.
\ee
These formulas also show that a small anisotropy induces sudden changes
of the edge states in a weak magnetic field:
When $q$ increases for a fixed $t>1$, 
$N_{\rm zigzag}$ increases toward $2\pi$ and $N_{\rm bearded}$ 
reaches zero at $q=\ln 2/\ln t$.
And for a fixed $t<1$, 
$N_{\rm zigzag}$ goes to zero and
$N_{\rm bearded}$ increases toward $2\pi$ 
as $q$ increases.
Thus in a weak magnetic field ($q\gg 1$),
there are abrupt changes in $N_{\rm zigzag}$ and $N_{\rm bearded}$ at $t=1$.

\begin{figure}
 \begin{center}
\includegraphics[width=8.0cm,clip]{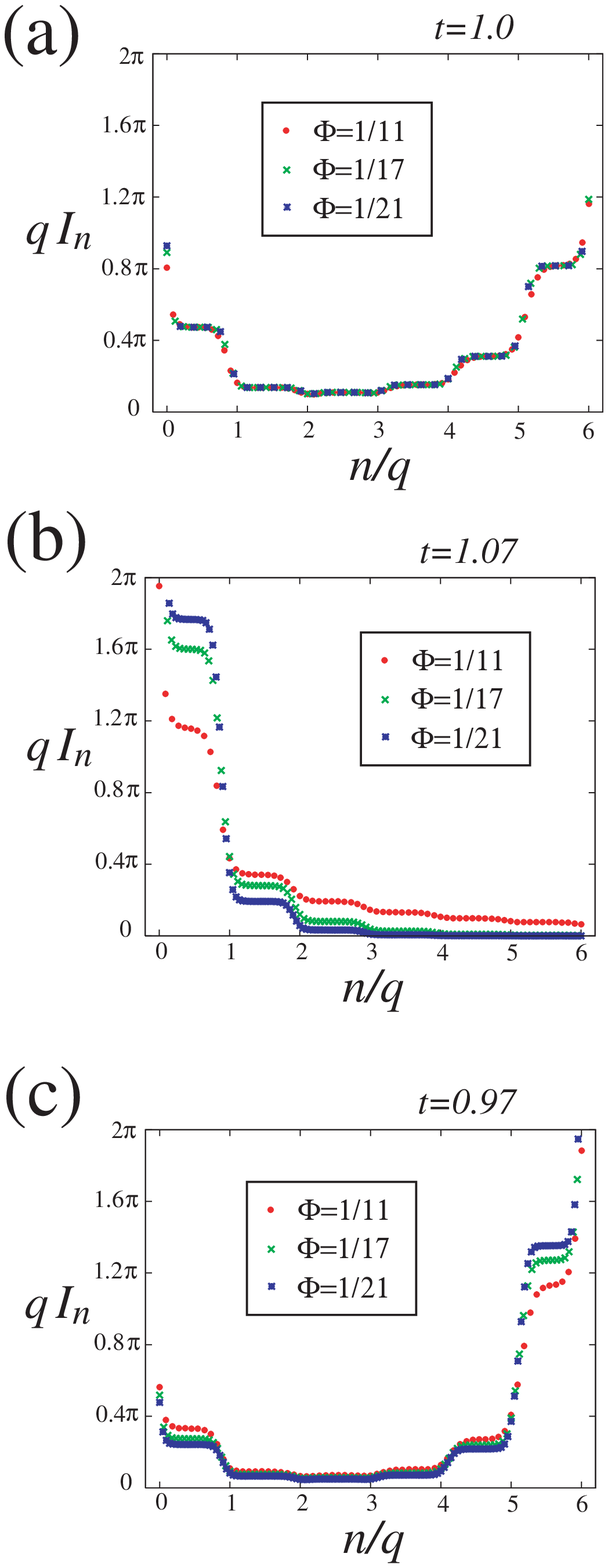}
\caption{\label{fig:zigzag_bearded_STM}(Color online)
 $q I_n$ as a function of $n/q$ for 
$\Phi=1/11$, $1/17$, and $1/21$.
(a) $t=1.0$, (b) $t=1.07$, (c) $t=0.97$.}
\end{center}
\end{figure}

\begin{figure}
 \begin{center}
\includegraphics[width=8.0cm,clip]{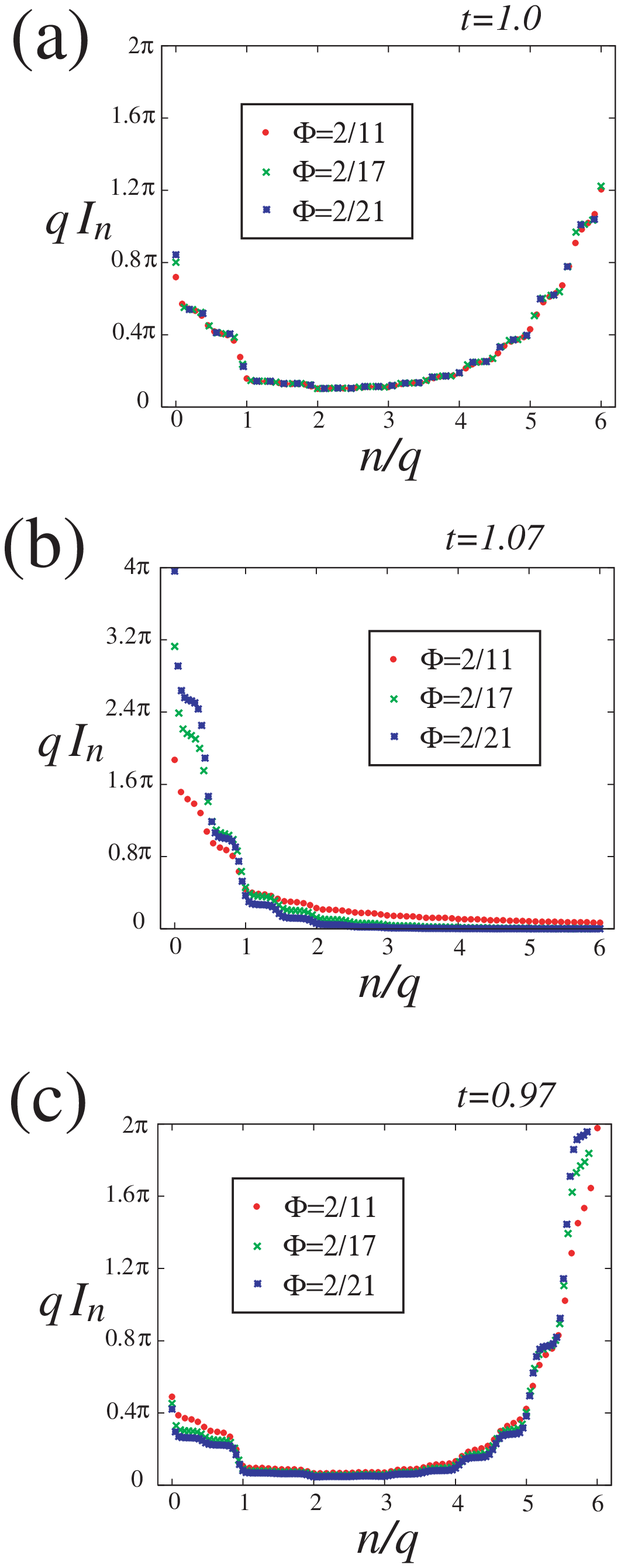}
\caption{\label{fig:zigzag_bearded_STM2}(Color online)
 $q I_n$ as a function of $n/q$ for 
$\Phi=2/11$, $2/17$, and $2/21$.
(a) $t=1.0$, (b) $t=1.07$, (c) $t=0.97$.
}
\end{center}
\end{figure}

In Figs.\ref{fig:zigzag_bearded_STM} and \ref{fig:zigzag_bearded_STM2}, 
we show scaled plots of $I_n$ as a function of $n/q$.
Here we have taken $L_x=6 q$.
Because of the normalization condition (\ref{normalize}), $I_n$ decreases as
$\sim 1/L_x$ when
$L_x$ increases.
To remove the artificial dependence on $L_x$, we plot $qI_n$ 
instead of $I_n$.  
For $t=1$, $qI_n$'s for different $\Phi(=p/q)$'s with the same $p$ 
fall on a common curve\cite{MA08_2}. 
However, for $t>1$, $q I_n$ comes to be localized on $q$ sites 
from the zigzag edges as $\Phi$ (with the same $p$) decreases.
In contrast, for $t<1$, $q I_n$ comes to be localized on $q$ sites 
from the bearded edges as $\Phi$ (with the same $p$) decreases.
In Table \ref{STM_1}, we also compare numerical data
and analytical results presented above.
With relative discrepancies less than $10\%$,
they exhibit good agreements.

\begin{figure}
 \begin{center}
 \includegraphics[width=7.0cm,clip]{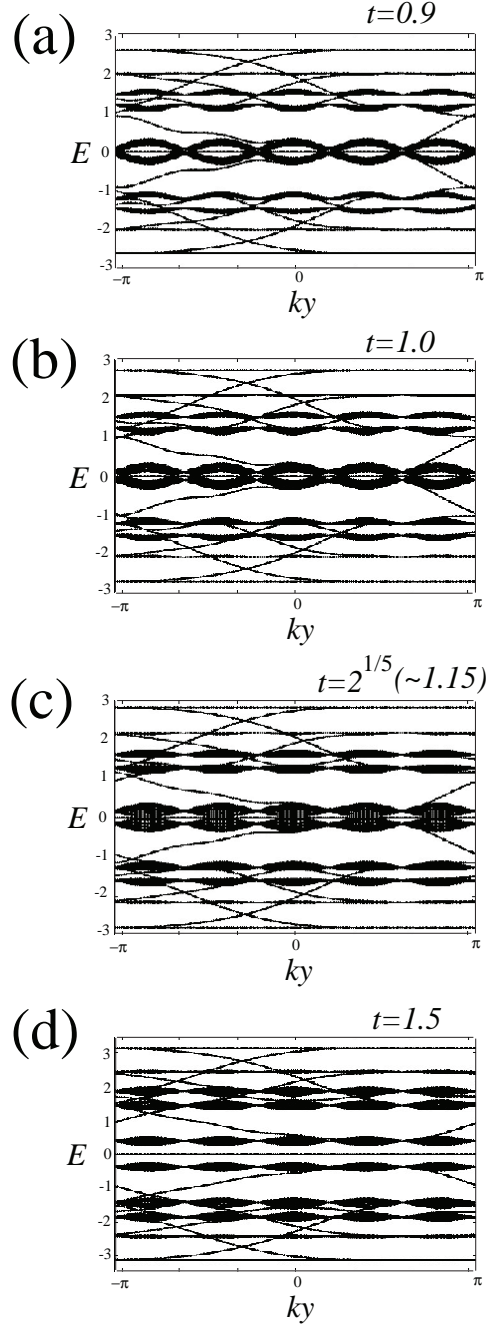}
\caption{\label{fig:zigzag_bearded_q=5000}
Energy spectra of honeycomb lattices with zigzag and bearded edges
for $\Phi=1001/5000(\sim 1/5)$ and $L_x=50$.
(a) $t=0.9$, (b) $t=1.0$, (c) $t=2^{1/5}(\sim 1.15)$ (d) $t=1.5$.}
\end{center}
\end{figure}

\begin{figure}
 \begin{center}
 \includegraphics[width=8.0cm,clip]{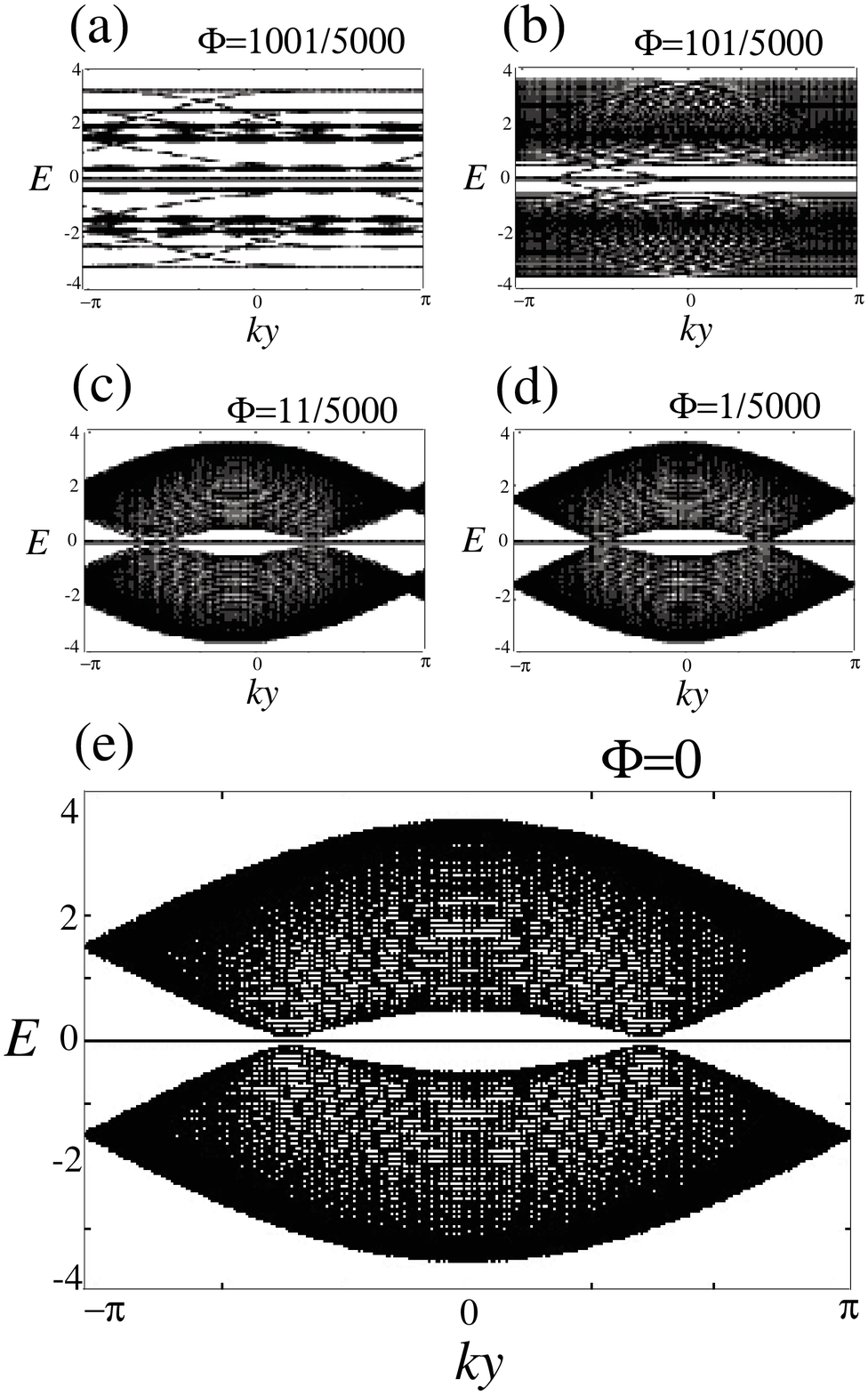}
\caption{\label{fig:zigzag_bearded_small}
Energy spectra of honeycomb lattices with zigzag and bearded edges
for $t=1.5$ and $L_x=50$. (a) $\Phi=1001/5000$, (b) $\Phi=101/5000$,
(c) $\Phi=11/5000$, (d) $\Phi=1/5000$.
We also show energy bands in the absence of a magnetic field in (e).}
\end{center}
\end{figure}

So far, we have assumed that $L_x \gg q$.
When $L_x \lesssim q$, the above arguments cannot be justified.
However, numerical calculations suggest that if $L_x\gg 1/\Phi$ ($p L_x
\gg q$), the particular edge states presented above appear again.
In Fig.\ref{fig:zigzag_bearded_q=5000}, 
we show the energy bands for $L_x=50$ and $\Phi=1001/5000 (\sim 1/5)$,
which are found to be indistinguishable from those for 
$\Phi=1/5$ (Fig.\ref{fig:zigzag_bearded_q=5}).
It is also found that if the magnetic field decreases
and $L_x \lesssim 1/\Phi$ is realized,
then the energy bands approach those without a magnetic field,
which is illustrated in Fig.\ref{fig:zigzag_bearded_small}.


\begin{table}
\begin{center}
\caption{Numerical data, $N_{\rm zigzag}$, $N_{\rm bearded}$,
 and analytical results, $N_{\rm zigzag}^{(\rm A)}$,
 $N_{\rm bearded}^{(\rm A)}$.
(a) $t=1.0$ (b) $t=1.07$ (c) $t=0.97$.
The relative discrepancies between numerical data and 
analytical results, 
$\delta_{\rm zigzag}=|(N_{\rm zigzag}^{(\rm A)}-N_{\rm zigzag})/N_{\rm zigzag}|$ and $\delta_{\rm bearded}=|(N_{\rm bearded}^{(\rm A)}-N_{\rm bearded})/N_{\rm bearded}|$, are also shown.}
\label{STM_1}
\begin{tabular}{|c||c|c||c|c||c|c|}
\multicolumn{7}{c}{}\\
\multicolumn{7}{c}{ {\large (a)} {\large$t=1.0$} } \\
\hline
$\Phi$ &  $N_{\rm zigzag}$ & $N_{\rm bearded}$ & $N_{\rm zigzag}^{(\rm A)}$ & $N_{\rm bearded}^{(\rm A)} $ &  $\delta_{\rm zigzag}$ & $\delta_{\rm bearded}$ \\
\hline\hline
1/11  & 1.484 & 2.562 &  & & 0.0770 & 0.0411\\
\cline{1-3}\cline{6-7}
2/11  & 1.484 & 2.562 &  & & 0.0770 & 0.0411\\
\cline{1-3}\cline{6-7}
1/17 & 1.485 & 2.563 & $2\sqrt{3}-\frac{2}{3}\pi$ & $\frac{4}{3}\pi-\sqrt{3}$ & 0.0776 & 0.0415\\
\cline{1-3}\cline{6-7}
2/17 & 1.485 & 2.563 &  &  & 0.0776 & 0.0415\\
\cline{1-3}\cline{6-7}
1/21 & 1.486 & 2.563 &   & & 0.0783 & 0.0415\\
\cline{1-3}\cline{6-7}
2/21 & 1.486 & 2.563 &   & & 0.0783 & 0.0415\\
\hline
\end{tabular}

\begin{tabular}{|c||c|c||c|c||c|c|}
\multicolumn{7}{c}{}\\
\multicolumn{7}{c}{ {\large (b)} {\large $t=1.07$} } \\
\hline
$\Phi$ &  $N_{\rm zigzag}$   & $N_{\rm bearded}$ & $N_{\rm zigzag}^{(\rm A)}$ & $N_{\rm bearded}^{(\rm A)}$ &  $\delta_{\rm zigzag}$ 
& $\delta_{\rm bearded}$ \\
\hline\hline
1/11  & 3.635 & 0.2336 & 3.447  &   & 0.0517 & \\
\cline{1-3} \cline{6-6}
2/11  & 3.635 & 0.2335 &   &   & 0.0517 & \\
\cline{1-4} \cline{6-6}
1/17 & 5.027 & 0.009609 & 5.024 & 0  & 0.000597 & -\\
\cline{1-3} \cline{6-6}
2/17 & 5.027 & 0.009591 &  &   & 0.000597 & \\
\cline{1-4} \cline{6-6}
1/21 & 5.550 & 0.0007872 & 5.550  &  & 0.00 & \\
\cline{1-3} \cline{6-6}
2/21 & 5.550 & 0.0007847 &  &  & 0.00 & \\
\hline
\end{tabular}

\begin{tabular}{|c||c|c||c|c||c|c|}
\multicolumn{7}{c}{}\\
\multicolumn{7}{c}{ {\large (c)} {\large $t=0.97$} } \\
\hline
$\Phi$ & $N_{\rm zigzag}$  & $N_{\rm bearded}$ & $N_{\rm zigzag}^{(\rm A)}$ & $N_{\rm bearded}^{(\rm A)}$ &  $\delta_{\rm zigzag}$ & $\delta_{\rm bearded}$ \\
\hline\hline
1/11  & 1.040 & 3.553 & 0.9665  & 3.484 & 0.0707 & 0.0194\\
\cline{1-3} \cline{6-7}
2/11  & 1.040 & 3.553 &   &  & 0.0707 & 0.0194\\
\hline
1/17 & 0.8618 & 3.992 & 0.8017 & 3.936  & 0.0697 & 0.0140\\
\cline{1-3} \cline{6-7}
2/17 & 0.8618 & 3.992 &  &   & 0.0697 & 0.0140\\
\hline
1/21 & 0.7610 & 4.248 & 0.7083  & 4.198 & 0.0693 & 0.0118\\
\cline{1-3} \cline{6-7}
2/21 & 0.7610 & 4.248 &   &  & 0.0693 & 0.0118\\
\hline
\end{tabular}

\end{center}
\end{table}

\section{Summary and discussion}
\label{sec:summary}

In this paper, we examined behavior of zero modes 
and a gap around zero energy 
for a tight-binding model on the anisotropic honeycomb lattice
in a magnetic field, whose anisotropy is controlled 
by the hopping parameter $t$.
It was found that zero modes exist 
for all (rational) $\Phi$ for $0<t\le 1$, and a gap around zero energy 
opens by a tiny anisotropy for the graphene in a magnetic field.
This is contrasted with the case for the square lattice,
where zero modes always exist for all (rational) $\Phi$ 
when we change the ratio of the hopping parameters $t_x/t_y$\cite{MK89}.
For $1<t<2$, a gap around zero energy in a weak magnetic field 
behaves as a non-perturbative and exponential form 
as a function of the magnetic field.
This non-analytic behavior is naturally explained by tunneling effects 
between energy levels around two Dirac zero modes 
in the absence of a magnetic field.
At $t=2$, the gap around zero energy in a weak magnetic field
makes a transition from an exponential (non-perturbative) 
to a power-law (perturbative) behavior 
as a function of the magnetic field.
In particular, an explicit form of the gap around zero energy
near the transition point is obtained by the WKB method.
For $t>2$, energy bands in a weak magnetic field 
show linear dependence on a magnetic field.

We also examined edge states with zero energy.
The condition for the existence of zero energy edge states in a magnetic field
is analytically derived. On the basis of the condition, 
it is found that the anisotropy of the hopping integrals induces 
abrupt changes of the number of zero energy edge states,
which depend on the shapes of the edges sensitively.

Finally, we would like to discuss possible experimental realization of
anisotropy of the hopping integrals.
Recently, it was experimentally found that a reversible and controlled 
uniaxial strain can be produced in graphene\cite{ZH08,ZH08_2}.
Therefore, we can expect that a small anisotropy of the hopping integrals 
is realized by the uniaxial strain.
On the other hand, in order to realize a large anisotropy such as $t\sim 2$, 
cold atoms in optical honeycomb lattices created by laser beams 
would serve as alternatives\cite{PD08}.
In these lattices, a large anisotropy could be induced and controlled
by changing the intensities of the laser fields\cite{SZ07}. 
In addition, by using Raman processes induced by laser fields,
effective magnetic fields can be generated in the optical lattice\cite{DJ03}. 


{\it Note added.}
After submission of this paper, we became aware of recent independent work
which has some overlap with ours for $t\sim 2$\cite{GM09,GM092}.
We are grateful to G. Montambaux for pointing out these papers.

\begin{acknowledgments}

This work was supported in part by Global COE Program
``the Physical Sciences Frontier,'' MEXT, Japan for K. E.
This work was also supported in part by NSF grant DMR-05-41988
for B. I. Halperin.

\end{acknowledgments}

\appendix

\section{Tight-binding model on a honeycomb lattice with a variable
lattice spacing}
\label{sec:appendixA}
\begin{figure}
 \begin{center}
  \includegraphics[width=8.0cm,clip]{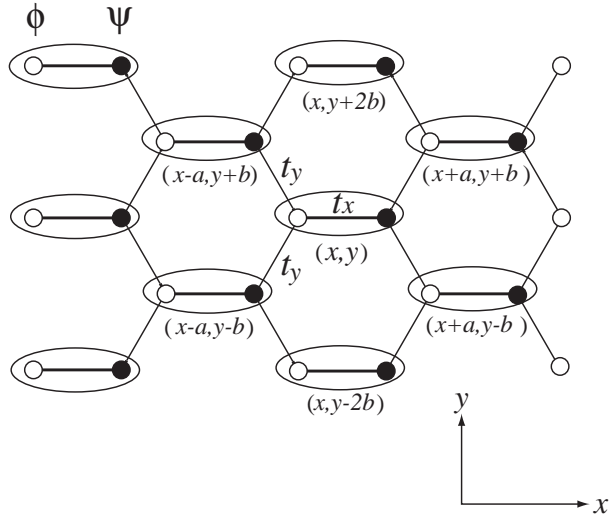}
\caption{\label{fig:honeycomb_lattice_scale}
 A honeycomb lattice with a variable lattice spacing. 
 If $a=\sqrt{3}b$, we have the hexagonal symmetry
 for each sublattice.
}
\end{center}
\end{figure} 

In this appendix, we consider 
a honeycomb lattice with a variable lattice spacing 
shown in Fig.\ref{fig:honeycomb_lattice_scale}.
The tight-binding equation in a magnetic field is given by
\be
E\psi(x,y)&=& t_y[ \phi(x+a,y-b)+e^{2 i\pi\frac{\Phi}{a} x}\phi(x+a,y+b)]
+t_x\phi(x,y), \nonumber\\
E\phi(x,y)&=&t_y[\psi(x-a,y+b)+e^{-2 i\pi\frac{\Phi}{a} (x-a)}\psi(x-a,y-b)]
+t_x\psi(x,y),
\label{magnetic_scale}
\ee
where a magnetic flux through a unit hexagon 
with an area of $S=2ab$ is given by $2\pi\Phi$. 
Let us write 
\be
x=na,\quad y=mb,
\ee
then (\ref{magnetic_scale}) gives
\be
E\psi(na,mb)&=& t_y[\phi((n+1)a,(m-1)b)
+e^{2 i\pi\Phi n}\phi((n+1)a,(m+1)b)]+t_x\phi(na,mb), \nonumber\\
E\phi(na,mb)&=&t_y[\psi((n-1)a,(m+1)b)+e^{-2 i\pi\Phi (n-1)}
\psi((n-1)a,(m-1)b)]+t_x\psi(na,mb).~~~~~~
\label{magnetic_scale2}
\ee
Then, by writing $\psi_{n,m}=\psi(na,mb)$, $\phi_{n,m}=\phi(na,mb)$,
and putting $t=t_x/t_y$, $t_y=1$,
we obtain the tight-binding equation (\ref{magnetic2}) 
from (\ref{magnetic_scale2}).

When we have lattice spacings $a$ and $b$ in the $x$ and $y$ directions,
respectively, as shown in Fig.\ref{fig:honeycomb_lattice_scale}, 
the momenta $q_x$ and $q_y$ in the $x$ and $y$ directions 
are related to $k_x$ and $k_y$ defined in (\ref{Bloch_uv}) as
\be
k_x=a q_x, \quad k_y=b q_y.
\label{kq1}
\ee
Especially, for $a=\sqrt{3} b$, where each sublattice has
the hexagonal symmetry, we have
\be
k_x=a q_x, \quad k_y=\frac{a}{\sqrt{3}} q_y.
\label{kq2}
\ee

Here, we mention the relation between
the magnetic field $B$ and the magnetic flux $\Phi$. 
They are related as
\be
BS=2ab B=2\pi\Phi,
\ee
that is,
\be
B=\frac{\pi\Phi}{ab}.
\label{B_Phi}
\ee
The tight-binding equation (\ref{magnetic2})
corresponds to $a=b=1$,
thus  from (\ref{B_Phi}) $B$ and $\Phi$ are related as
\be
B=\pi\Phi.
\ee

\end{document}